\documentclass[twocolumn,showpacs,preprintnumbers,amsmath,amssymb]{revtex4}

\input psfig.sty

\begin{document}

\title{Relativity, nonextensivity, and extended power law distributions}
\author{R. Silva$^{1,2}$\thanks{e-mail: rsilva@on.br, rsilva@uern.br} and
J. A. S. Lima$^{3}$\thanks{e-mail: limajas@astro.iag.usp.br}}
\smallskip
\address{$^{1}$Observat\'orio Nacional, Rua Gal. Jos\'e Cristino 77,
20921-400 Rio de Janeiro - RJ, Brazil}
\address{$^{2}$Universidade do Estado do Rio Grande do Norte,
59610-210, Mossor\'o, RN, Brazil}
\address{$^{3}$Departamento de Astronomia - IAG, Universidade de
S\~ao Paulo, 05508, S\~ao Paulo, SP, Brazil}
\date{\today}

\date{\today}

\begin{abstract}
A proof of the relativistic $H$-theorem by including nonextensive
effects is given. As it happens in the nonrelativistic limit, the
molecular chaos hypothesis advanced by Boltzmann does not remain
valid, and the second law of thermodynamics combined with a
duality transformation implies that the q-parameter lies on the
interval [0,2]. It is also proved that the collisional
equilibrium states (null entropy source term) are described by the
relativistic $q$-power law extension of the exponential Juttner
distribution which reduces, in the nonrelativistic domain, to the
Tsallis power law function. As a simple illustration of the basic
approach, we derive the relativistic nonextensive equilibrium
distribution for a dilute charged gas under the action of an
electromagnetic field $F^{{\mu\nu}}$. Such results reduce to the
standard ones in the extensive limit, thereby showing that the
nonextensive entropic framework can be harmonized with the
space-time ideas contained in the special relativity theory.

\end{abstract}
\pacs{05.90.+m; 05.20.-y; 03.30.+p; 05.70.Ln}

\maketitle

In the last few years, a great deal of attention has been paid to
the nonextensive Tsallis entropy both from theoretical and
observational viewpoints \cite{T88,ZA95,R96,SL99}. Recent
applications of the nonextensive entropy to an increasing number
of physical problems is beginning to provide a more definite
picture on the kind of scenarios where the new formalism proves to
be extremely useful [3-13].

At present, self-gravitating systems and plasma physics offer the
best framework for searching to nonextensive effects. The first
one is characterized by very strange kinetic and thermal
properties (see \cite{PP93} for recent publications on this
topic). Actually, collisionless stellar systems like galaxies are
endowed with negative specific heat, and the simplest density
profiles based on the Maxwellian distribution lead to infinite
mass (the so-called singular isothermal sphere). In the case of
plasmas, Boghosian's treatment for a two dimensional pure electron
plasma yielded the first experimental confirmation of Tsallis
theory \cite{BO96}, whereas experiments related to dispersion
relations for electrostatic plane-wave propagation also points to
a class of power law Tsallis velocity distributions \cite{LSS00}.
In reality, it is now widely believed that the nonequilibrium
properties of such systems away from Boltzmann-Gibbs state are not
completely understood \cite{TS04}. This nonextensive statistical
formalism also proved to be an useful construct for the analysis
of many interesting properties of linear and nonlinear
Fokker-Planck equations \cite{FokkerP}.

On the other hand, most of the observational or experimental
evidence supporting Tsallis proposal are related to the power-law
velocity distribution associated with Tsallis thermostatistical
description of the classical $N$-body problem \cite{SPL98}. For a
dilute gas of massive point particles, the nonextensive effects
are simply parameterized by the local entropy density formula
\begin{equation} \label{eq:1}
S_q=-k_{B}\int f^q\ln_q f d^{3}p ,
\end{equation}
where $k_{B}$ is the Boltzmann constant, $f$ is the distribution
function, $q$ is the nonextensive parameter and the
$q$-logarithmic function is defined by
\begin{equation}\label{eq:224}
\ln_q f\, =\, (1-q)^{-1}(f^{1-q}-1), \,\,\,\,\,\,\,\,\, (f>0)
\end{equation}
which recovers the standard Boltzmann-Gibbs entropy $S = -
k_{B}\int f \ln f {d^{3}p}$ in the limit $q\rightarrow 1$. In the
nonrelativistic limit, the time evolution of $S_q$ was analyzed
with basis on the Liouville and Fokker-Planck equations
\cite{BP98}, as well as through a nonextensive generalization of
the nonrelativistic Boltzmann $H$-theorem \cite{LSP01,Fernando04}.

As first recognized by Lima et al. \cite{LSP01} (hereafter paper
I), the attempts for extending Boltzmann kinetic theory by
including nonextensive effects, which basically means a
$q$-transport equation and the associated $H$-theorem, required a
departure from the celebrated molecular chaos hypothesis first
advanced by Boltzmann. In this connection, it is worth notice
that the $q$-Boltzmann equation of paper I differs from the one
proposed by Kaniadakis \cite{kania2001}, an approach based on the
kinetic interaction principle, only by the assumed form of the
collision integral.

Theoretically, beyond the applications closely related to the
nontrivial solutions of the nonrelativistic q-transport equation
and the associated transport coefficients \cite{LSS00}, it is
clearly necessary to go one step further, by extending the proof
of $H$-theorem to the relativistic and quantum domains. The basic
reasons are very well known, and have partially guided the
development of modern physics \cite{degrrot}. Actually, in the
case of the Maxwell-Boltzmann distribution, this basic program has
already been performed in detail to special relativity
\cite{juttner}, quantum theory \cite{TO}, as well as, by including
the gravitational interaction, to the context of general
relativity theory \cite{CH62}. In particular, the collisional
equilibrium of a relativistic gas of massive point particles is
described by the Juttner distribution function which contains the
number density, the temperature, and the local 4-momentum as free
parameters \cite{degrrot,juttner}.

In this letter, the nonrelativistic q-Boltzmann equation and the
$H$-theorem discussed in paper I are extended to the special
relativistic domain through a manifestly covariant approach. As we
shall see, the whole argument follows from a direct generalization
of the molecular chaos hypothesis and the expression of the
4-entropy flux in the spirit of the nonextensive Tsallis
prescription. The leitmotiv of this article is to shown that the
kinetic nonextensive approach can also be harmonized with the
space-time ideas contained in the special relativity theory.

To begin with, we recall that the proof of the standard
relativistic $H$-theorem is also based on the molecular chaos
hypothesis (Stosszahlansatz), i.e., the assumption that any two
colliding particles are uncorrelated. This means that the two
point correlation function of the colliding particles can be
factorized
\begin{equation}\label{Boltzmann2}
f(x, p, p_1) = f(x, p) f(x, p_1),
\end{equation}
or, equivalently,
\begin{equation}\label{Boltzmann2a}
\ln f(x, p, p_1) = \ln f(x, p) + \ln f(x, p_1),
\end{equation}
where the particles have 4-momentum $p\equiv p^\mu=(E/c,\bf{p})$
in each point $x \equiv x^\mu=(c t,\bf{r})$ of the space-time,
with their energy satisfying $E/c=\sqrt{{\bf p}^2+m^2 c^2}$ (in
the above expressions, $p$ and $p_1$, are the 4-momenta just
before collision). In what follows, we shown that the relativistic
nonextensive entropic measure is consistent with a slight
departing from ``Stosszahlansatz" (molecular chaos) when exact
correlations are introduced. Operationally, this means that one
must replace the logarithm functions appearing in
(\ref{Boltzmann2a}) by their nonextensive counterpart which are
represented by the q-logarithmic (power laws) defined by
(\ref{eq:224}). It should be recalled that the validity of the
chaos molecular hypothesis still remains as a very controversial
issue \cite{Zeh}. Probably, the unique consensus is that it is by
no means a consequence of the laws of mechanics, and, as shown in
paper I, the ``Stosszahlansatz" is not responsible by the
irreversible content of the Boltzmann approach.

Let us now consider a relativistic rarified gas containing $N$
point particles of mass $m$ enclosed in a volume $V$, and under
the action of an external 4-force field $F^\mu$. From a kinetic
viewpoint, the states of the gas must be characterized by a
Lorentz invariant one-particle distribution function $f(x,p)$. By
definition, the quantity $f(x,p) d^3xd^3p$ gives, at each time
$t$, the number of particles in the volume element $d^3xd^3p$
around the particles space-time position $x$ and momentum ${\bf
p}$. By taking into account the nonrelativistic treatment (see
\cite{LSP01}), one may assume that the temporal evolution of the
relativistic distribution function $f(x,p)$ is driven by the
following q-transport equation
\begin{equation}\label{relBot}
p^{\mu}\partial_\mu f+m F^\mu{\partial{f}\over\partial
p^{\mu}}=C_q(f),
\end{equation}
where the index $\mu$ take the four values 0,1,2,3, while
$\partial_\mu=(c^{-1}\partial_t,\nabla)$ indicates differentiation
with respect to time and space coordinates, respectively, and
$C_q$ denotes the relativistic $q$-collisional term. Note that the
left-hand-side of (\ref{relBot}) is just the total derivative of
the distribution function or the ``streaming term". This means
that the nonextensive effects can be manifested only through the
collisional term which is a local slowly varying function of
$f(x,p)$. The collision integral, $C_q(f)$, must be consistent
with the energy, momentum, and the particle number conservation
laws, and its specific structure must be such that the standard
result is recovered in the limit $q\rightarrow 1$. At this
point, it is interesting to compare the approach developed here
which is based on equation (\ref{relBot}) with the one proposed by
Lavagno \cite{lavagno2}. In the latter work, all the nonextensive
effects are quantified by assuming a modified Boltzmann equation
to the quantity $f^{q}$ (see Eq. (13) of the quoted paper). In
particular, this means that such theories must lead to different
predictions of the physical quantities, as for instance, the
expressions for the transport coefficients.

Now, since $C_q(f)$ leads to a nonnegative local $q$-entropy
source, that is, $\tau_q(x)\equiv
\partial_\mu S^\mu_q$, where $S^{\mu}$ is the 4-entropy flux
(an identically vanishing quantity for equilibrium states), its
general form reads
\begin{equation}
C_q(f)= {c\over 2} \int F \sigma R_q (f,f'){d^3 p_1\over E_1}
d\Omega,
\end{equation}
where $d\Omega$ is an element of the collision solid angle, the
scalar $F$ is the invariant flux, which is equal to
$F=\sqrt{(p_\mu p^\mu_1)^2 -m^4 c^4}$, and $\sigma$ is the
differential cross section of the collision $p + p_1\rightarrow p'
+ p'_1$ (see Ref \cite{degrrot} for more details). All quantities
are defined in the centre-of-mass system of the colliding
particles. In a point of fact, relativity enters only in the
definition of $F$, and implicitly through the differential
cross-section $\sigma$. The quantity $R_q(f,f')$ is a difference
of two correlation functions which are assumed to satisfy a
$q$-generalized form of the molecular chaos hypothesis expressed
as \cite{LSP01}
\begin{eqnarray*} R_q(f,f') = e_q({f'}^{q-1}\ln_q
f'+{f'}_1^{q-1}\ln_q f'_1)
\end{eqnarray*} \begin{equation} -e_q(f^{q-1}\ln_q f+
f_1^{q-1}\ln_q f_1),
\end{equation}
where primes refer to the distribution function after collision.
Note that in the limit $q \rightarrow 1$ the above expression
reduces to $R_1=f' f'_1-f f_1$, thereby showing that the molecular
chaos hypothesis is readily recovered. Similarly, the nonextensive
4-entropy flux reads
\begin{equation}\label{qflow}
S^\mu_q=-k_{B} c^{2}\int p^\mu f^q\ln_q f {d^3 p\over E},
\end{equation}
and as should be expected, $c^{-1}S_q^{0}$ is just the local
Tsallis' entropy density as given by (\ref{eq:1}). Now, in order
to obtain the source term, we first take the 4-divergence of
$S_{q}^{\mu}$
\begin{equation}
\partial_\mu S^\mu_q \equiv \tau_q= -k_{B}c^{2} \int(qf^{q-1}\ln_q f+1)
p^\mu\partial_\mu f {d^3p\over E},
\end{equation}
and combining with the nonextensive relativistic Boltzmann
equation (\ref{relBot}), one may rewrite the above expression in
the following form
\begin{equation}\label{BoltzmannRel}
\tau_q = -{k_{B}c^3\over 2} \int F \sigma(qf^{q-1}\ln_q f+1)R_q
{d^3p\over E}{d^3p_1\over E_1}d\Omega.
\end{equation}
At this point, it is convenient to rewrite $\tau_q$ in a more
symmetrical form by using some elementary symmetry operations
which also take into account the inverse collisions. First we
notice that by interchanging $p$ and $p_1$ the value of the
integral is preserved. This happens because the scattering cross
section and the magnitude of the flux are invariants
\cite{degrrot}. In addition, the value of $\tau_q$ is not
altered if we integrate with respect to the variables $p'$ and
$p'_1$. Actually, although changing the sign of $R_q$ in this step
(inverse collision), the quantity ${d^3pd^3p_1/p^0 p^0_1}$ is also
a collisional invariant \cite{degrrot}. Finally, as we have done
in paper I, we apply a ``duality" transformation (see discussion
below and Ref. \cite{karlin}) of the form $f^{q-1}\ln_q f=\ln_{q*}
f$, where the new nonextensive parameter is related to the old one
by $q*=2-q$. As one may check, such considerations imply that the
$q$-entropy source term can be written as
\begin{eqnarray*}\label{ST}
\tau_q(x)={q k_{B}
c^3\over 8}\int F\sigma (\ln_{q^*} f' + \ln_{q^*} f'_1 - \ln_{q^*}
f -\ln_{q^*} f_1)
\end{eqnarray*}
\begin{equation}
[(e_q(\ln_{q^*} f' + \ln_{q^*} f'_1)-e_q(\ln_{q^*} f -\ln_{q^*}
f_1)]{d^3p\over E}{d^3p_1\over E_1} d\Omega.
\end{equation}
This is our main result, and the reader should compare it with the
nonrelativistic expression deduced in paper I. As widely known,
the irreversible nature of thermodynamics emerging from molecular
collisions is recovered if the above quantity is positive
definite. In the present case, such a condition can be guaranteed
in two steps. First, we notice that the integrand of
\begin{eqnarray*} (\ln_{q^*} f' +
\ln_{q^*} f'_1 - \ln_{q^*} f -\ln_{q^*} f_1)\times
\end{eqnarray*}
\begin{equation}
[e_q(\ln_{q^*} f' + \ln_{q^*} f'_1)-e_q(\ln_{q^*} f -\ln_{q^*}
f_1)],
\end{equation}
is always positive for any pair of distributions $(f,f_1)$ and
$(f',f'_1)$. This means that the sign of the 4-entropy source is
now completely determined by the sign of the nonextensive
parameter. Therefore, if the second law is to be obeyed
\cite{degrrot,GP}, the values of this parameter must be restricted
to $q \geq 0$. In other words, when $q < 0$, the relativistic
q-entropy source of a given volume element decreases in the course
of time. Note that the boarder case ($q=0$) seems to be physically
meaningless, since the entropy is constant regardless of the
solution obtained from the transport equation with a non-null
collision integral. In this concern, one may ask if the $q$
parameter is limited from above. As one may check, repeating all
the calculations present until now with a duality transformation,
i.e., by taking $q*=2-q$ in Tsallis entropy, Eq. (\ref{eq:1}), it
is easy to conclude that $q*>0$. Therefore, the duality
transformation together with the relativistic $H$-theorem imply
that $q$ is constrained on the interval [0,2] (as first pointed
out by Karlin et al. \cite{karlin}, such a result is also valid in
nonrelativistic theory \cite{LSP01}). In particular, this means
that the upper bound of $q$, i.e. $q<2$ is not a purely quantum
mechanics restriction, as recently claimed in the literature
\cite{abe}.

In order to complete the proof of the theorem, let us now derive
the nonextensive Juttner distribution. Such a function is the
relativistic version of the q-power Tsallis distribution
\cite{SPL98,LSP01}, and must be obtained as a natural consequence
of the relativistic $H$-theorem. As happens in the classical
case, $\tau_q=0$ is a necessary and sufficient condition for local
and global equilibrium. Since the integrand appearing on the
expression of $\tau_q$ must be positive definite, this occurs if
and only if
\begin{equation}
\ln_{q^*} f' + \ln_{q^*} f'_1 = \ln_{q^*} f +\ln_{q^*} f_1,
\end{equation}
where the 4-momenta are connected through a conservation law
($p^\mu + p^\mu_1 = {p'}^\mu+{p'}^\mu_1$) which is valid for any
binary collision. Therefore, the above sum of $q$-logarithms
remains constant during a collision. It is a summational
invariant. In the relativistic case, the most general collision
invariant is a linear combination of a constant plus the
four-momentum $p^\mu$ \cite{degrrot}. Consequently, we must have
\begin{equation}\label{joia}
\ln_{q^*} f^0 (x,p)=\alpha(x)+\beta_\mu p^\mu,
\end{equation}
where $\alpha(x)$ is a scalar, $\beta_\mu$ a 4-vector, and $p^\mu$
is the four-momentum. After simple algebra, we may rewrite
(\ref{joia}) as a relativistic nonextensive distribution
\begin{equation}\label{MBJgeneralized}
f^0(x,p)=[1-(1-q)(\alpha(x)+\beta_\mu p^\mu)]^{1/1-q},
\end{equation}
with arbitrary space and time-dependent parameters $\alpha(x)$ and
$\beta_\mu(x)$. The above expression is the relativistic version
of the $q$-Tsallis distribution \cite{LSP01}. The function $f^0
(x,p)$ is the most general expression which leads to a vanishing
collision term and entropy production, and reduces to Juttner
distribution in the limit ${q \rightarrow 1} $. However, it is not
true in general that $f^0(x,p)$ is a solution of the transport
equation. This happens only if $f^0$ also makes the left-hand-side
of the transport equation (\ref{relBot}) to be identically null.
Nevertheless, since (\ref{MBJgeneralized}) is a power law, the
transport equation implies that the parameters $\alpha(x)$ and
$\beta_\mu (x)$ must only satisfy the constraint equation
\begin{equation}
p^\mu\partial_\mu \alpha(x)+p^\mu p^\nu\partial_\mu
\beta_\nu(x)+m\beta_\mu(x) F^\mu(x,p)=0.
\end{equation}
The nonextensive distribution of the form (\ref{MBJgeneralized}),
with the specific parameters obeying the above equation, describes
the relativistic (nonextensive) local equilibrium states.

For illustration purposes, let us now consider a relativistic gas
under the action of the Lorentz 4-force $F^\mu(x,p)=-(Q/mc) F^{\mu\nu}(x)p_\nu,$
where $Q$ is the charge of the particles and $F^{\mu\nu}$ is the
Maxwell electromagnetic tensor. Following standard lines, it is
easy to show that the local equilibrium function in the presence of an
external electromagnetic field reads
\begin{equation}
f(x,p)=\left[1-(1-q)\left({\mu-[p^\mu+c^{-1}Q A^\mu (x)]U_\mu\over
k_{B} T}\right)\right]^{1/1-q},
\end{equation}
where $U_\mu$ is the mean four-velocity of the gas, $T(x)$ is the
temperature field, $\mu$ is the Gibbs function per particles, and
$A^\mu(x)$ the four potential. Note that the above expression in
the limit $q \rightarrow 1$ reduces to the well known expression
\cite{degrrot,Hakim}
\begin{equation}
f(x,p)=\exp\left({\mu-[p^\mu+c^{-1}Q A^\mu (x)]U_\mu\over k_{B}
T}\right).
\end{equation}

Summarizing, we have proposed a $q$-generalization of the
relativistic Boltzmann's equation and the associated
$H$-theorem along the lines of Tsallis' nonextensive kinetic
theory. We have found that the nonextensive ideas can be
consistently extended in order to incorporate the space-time
concepts of the special relativity. In addition, since the basic
results were derived in a manifestly covariant way, their
generalization to the general relativistic framework can be
readily accomplished.

It is worth notice that the relativistic counterpart of the
$H$-theorem constrains the physically allowed values for the
$q$-parameter (as it occurs in the Newtonian regime), and its
proof also does not require the ``Stosszahlansatz" (molecular
chaos) Boltzmann assumption. By the reasons discussed before, the
$q$-nonextensive contributions must appear explicitly only in the
collisional term of the q-transport equation, and, as such, the
approach followed here differs profoundly from another attempts to
generalize the Boltzmann equation within the spirit of Tsallis'
framework \cite{lavagno,lavagno2}. As should be expected, the relativistic
class of $q$-distributions reduce to the standard Juttner result
in the extensive limit $q=1$. However, different from the
extensive Maxwell-Boltzmann-Juttner approach, correlations are
extremely relevant in the nonextensive context (see also paper I),
and, more important, the corresponding modifications in the
collisional term are consistent with the standard laws of
microscopic dynamics. As we have shown, such correlations are
exactly described and form the physical basis of the nonextensive
$H$-theorem either for the relativistic and nonrelativistic
regimes.

It should be stressed that the combination of the relativistic
$H$-theorem and duality transformation \cite{karlin} restricted
the $q$-parameter on the range, [0,2], which is exactly the same
result of the nonrelativistic quantum domain \cite{abe} and of the
consistent framework for generalized satistical mechanics
\cite{kaniadakis05}. It should be noticed, however, that the
allowed range of $q$ may be even smaller if one takes into account
the finite normalization condition and the negativeness of heat
capacity. In the nonrelativistic regime, for instance, it has been
shown that $q$ must be smaller than 5/3
\cite{PP93,hansen04,limrs04}. Finally, two points to be noted here
and explored in the near future are the possible connection
between the relativistic nonextensive function and the
kappa-distributions \cite{leubner04}, and the search to the
expressions of the q-relativistic transport coefficients
\cite{BSL03}.

{\it Acknowledgments:} The authors are grateful to Angel Plastino
and Jailson Alcaniz for helpful discussions. This work was partially supported by the
Conselho Nacional de Desenvolvimento Cient\'{\i}fico e
Tecnol\'ogico (CNPq - Brazil).

\end{document}